# AUTOMATED ENVIRONMENTAL MONITORING INTELLIGENT SYSTEM BASED ON COMPACT AUTONOMOUS ROBOTS FOR THE SEVASTOPOL BAY

Y.E. Shishkin[1,2], Aleksandr N. Grekov[1,*]

[1] Institute of Natural and Technical Systems, Russia, Sevastopol, Lenin St., 28
[2] Sevastopol State University, Russia, Sevastopol, Universitetskaya St., 33
*E-mail: oceanmhi@ya.ru

This paper proposes an intelligent system concept for automatic monitoring of aquatic environment main parameters in order to detect their anomalies and assess quantitative and qualitative indicators, including the determination of the field under study spatial and temporal characteristics. The system is built on the basis of a small autonomous surface robots network. A conceptual model of a monitoring system for the implementation of environmental parameters automated integrated monitoring throughout the entire observation field is proposed. A software model has been developed and simulation experiments have been conducted to calculate the main indicators and assess environment spatial and temporal variability. According to the results of the simulation, control maps of the stations optimal density were formed. The proposed approach to solving the problem of monitoring the aquatic environment in comparison with the traditional has such advantages as scalability, flexibility, speed of deployment and clotting, self-organization, the ability to create a wide field of view by changing the number of robots.

**Introduction**. There are many ways to obtain data to understand the complex dynamic processes occurring in the oceans [1]. Modern technologies allow large-scale observations to be carried out using satellites or other means of remote sensing, but there is still a huge demand for in situ measurements [2]. These measurements can be used to calibrate satellite data, in particular to correct vertical profiles, but their main purpose is to collect data on processes that occur on a much smaller scale. The need to control the dynamics of many processes occurring in the marine environment has always required more effective methods of increasing the sampling rate in the space-time domain. There is a huge variety of robotic systems that have been developed to provide such measurements, and it is not surprising that they are playing an increasingly important role in the measurement of the marine environment.

There are many known designs of autonomous surface vehicles (ASVs) capable of carrying meters to monitor the characteristics of the ocean and atmospheric boundary layers. The operation of such devices can last for several days or weeks, and information from them, if necessary, can be transmitted in real time. To collect data underwater, universal platforms are widely used: Autonomous Underwater Vehicles (AUVs), which allow research in vast areas with minimal operating costs [3]. In most applications, AUVs are typically programmed to follow predetermined paths while collecting the necessary data, providing efficient representations of 3D fields.

For scientific research, robotic systems are priority tools that allow one to determine the characteristics of the dynamic processes of the marine environment, and their efficiency is steadily increasing: the density of measurements in space and time, as well as the duration and scale of tasks being performed, increases. Such systems can operate around the clock in marine conditions with little or no human intervention. The increased onboard computing power gives these systems built-in intelligence to interpret data in real time and introduce new paradigms when collecting data in the marine environment. Robotic systems benefit even more from the simultaneous use of several vehicles for the same mission: they can work in cooperation, performing independent tasks or synchronous actions.

The general principles of functioning of automated systems based on autonomous vehicles and the problems associated with the implementation of such systems were considered earlier in [4]. Continuous environmental monitoring with the use of autonomous devices allows you to monitor the environmental situation in the sea areas in real time. In work [5], the main tasks that the regional automated system of environmental monitoring of sea areas are designed to solve are considered and the concept of its geographically distributed clusters, called information regions, is introduced, and the conceptual appearance of the system is synthesized and the basic principles of its construction are formulated.

However, a number of issues remain unresolved related to integrated environmental monitoring systems for specific water areas. In accordance with the methodological instructions of Roshydromet [6], for a comprehensive assessment of the state of surface waters, it is necessary to control 15 obligatory hydrochemical ingredients and water quality indicators. The optimal number of ingredients considered in the evaluation process can be from 10 to 25. In addition to hydrochemical indicators, observations should include determination of water salinity, water and air temperature, speed and direction of currents and winds, transparency and color of water [7], and observations of background characteristics, for example, of the Sevastopol Bay, should be carried out regularly. It is not possible to equip autonomous devices with such a number of measuring channels; moreover, not all hydrochemical parameters have in situ sensors. Therefore, it would be expedient to assemble autonomous devices with the main hydrological measuring channels and a channel that allows obtaining an integral toxicological characteristic regardless of the nature and composition of pollutants. For example, a channel based on a biomonitoring method using the behavioral responses of native species of aquatic organisms, and primarily bivalve molluscs, can be applied [8, 9].

To solve issues related to the environmental control of the Sevastopol Bay, the task is to monitor to detect anomalies in the quantitative indicators of the marine environment, characterized by heterogeneous parametric variability, in a certain limited area using an intelligent autonomous surface robot (IASR) or robot systems.

The problem is divided into two parts: the problem of anomaly recognition and the problem of the geometric location of the control points of the route of the robots. It should be noted that our study does not cover several critical components of robotic systems, such as Nested autonomy [10, 11], underwater navigation, communications and motion control [12–14], networks of underwater oceanographic devices [15].

On the one hand, it is necessary to define a mechanism for detecting anomalies using threshold, statistical or expert assessments (adaptive sampling), and on the other hand, to locate the points of the route (stations) in such a way that the detection is carried out with the highest quality (path planning). In order for the monitoring to be carried out in the most informative way, it is necessary that the step of the stations be coordinated with the variability of the indicators of the aquatic environment and the resolution of the meters.

The first problem is of a particular nature for each specific case and depends on the degree of spatial and temporal autocorrelation (variability) of the measured parameter, the resolution and reactivity of the measuring channels and the medium itself. This problem is best solved in the presence of a model of the environment that most fully describes the dynamics of all processes and allows assimilation of data in real time [16]. However, such models do not exist for all water areas, and their creation is a very time-consuming task that requires a huge array of pre-collected data.

A typical scenario of use includes the following stages: collection and structuring of monitoring data of the area of interest to the researcher over a long period; determination of a map of monitoring intervals, providing maximum information content for each station; survey of data with primary meters in accordance with the map; adaptive map correction of optimal intervals.

The classical problem of the optimal geometric location of route points (patrolling problem) is reduced to finding such a set of geometric route points that guarantee an effective solution to the monitoring problem. In the context of this article, the given initial conditions require redefining the very formulation of the problem, approaches to its solution and the concept of the effectiveness of solving the problem.

Due to the fact that the IASR is equipped with point meters of environmental indicators, its coverage (visibility) area at a point $(x, y)$ depends on the variability of the environment and is determined by the ratio

$$R_z(x, y) = \frac{r_{\max} \cdot \xi}{\sqrt{\left(\frac{\partial z}{\partial x}\right)^2 + \left(\frac{\partial z}{\partial y}\right)^2}}, \quad (1)$$

where $r_{\max}$ is the upper limit of the measuring range; $\xi$ - relative error of the measuring channel; $z = f_r(x, y)$ - the value of the measured parameter at point $(x, y)$, calculated using the regression equation of two variables.

Considering that the results generated by the model at this stage of its development will be used as input parameters in the route control system using the characteristics of wind speed and current, it is the formation of the field of optimal intervals for the location of stations that will be optimal, and not the search for specific points in the monitoring area. Using this field, it will be possible to find such an arrangement of stations that covers the entire monitoring area with a visibility zone and at the same time meets the criteria for a minimum for the number of stations, the total time of visiting all stations, energy consumption, taking into account the direction and speed of the wind, and other integral characteristics.

**Software modeling.** To test the developed algorithm, a software model was built, which allows a number of experiments to be carried out and the main indicators of the temporal and spatial variability of the environment to be calculated, and control maps of the optimal density of the station location to be generated.

A processed area of the terrain map containing the field of observation of the measured parameter corresponding to the monitoring area is received at the input of the program. Coverage areas for each robot are calculated based on the reference characteristics of the measuring channels and the gradient field of the medium. Obstacles and coastline do not affect coverage areas.

В результате анализа всех характеристик среды, поддающихся измерениями *in situ* As a result of the analysis of all the characteristics of the medium that can be measured in situ, the concentration of dissolved oxygen has the greatest variability and measurement accuracy. Let us consider the application of the algorithm on the map of the field of distribution of dissolved oxygen in the surface layer in the bay of Sevastopol, in the period from 1998 to 2007 [17], which has the form (Figure 1).

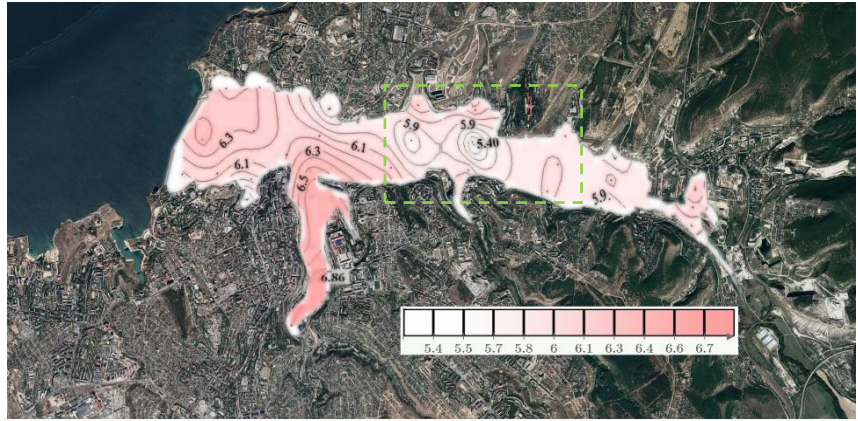

**Fig. 1.** Distribution of dissolved oxygen (ml / l) in the surface layer of the bay of Sevastopol, visualization was carried out according to the data of the atlas of oceanographic characteristics of the Sevastopol bay, MGI, 2010

A feature of gradient methods with a high sensitivity to changes in the parameters of the medium is the extremely high dependence of the quality of the solution on the step of the computational grid. With the same density of the grid defining the factor space, their error is an order of magnitude greater than when using the methods for estimating the absolute values of the field. Therefore, to simplify further calculations and improve their quality, we select the central region of the observation field (highlighted by a dotted line) and increase the density of the di-mensional grid. To do this, we describe a given set of points (Fig. 2 a) by two mathematical models specified in the form of a biquadratic regression equation, and the same, but with flexible threshold smoothing (Fig. 2 b, c).

It is assumed that the region $D \subset R_2$ and the rule defining the mapping $M(x, y) \in D \xrightarrow{f} w \in R_1$ are set, at $n$ arbitrarily specified points according to which each point $M$ is assigned the number $w$ according to rule $f$. Then the scalar field is expressed by the function $w = f(M) = f(x, y)$. The mathematical model in general form can be set in the form of rule $f_r$ such that satisfies condition (2), represented in the form

$$f(M(x, y)) = f_r(M(x, y))$$
$$\forall x \in X, y \in Y \quad , \tag{2}$$

whereby $f_r$ is defined for the entire $M(x, y) \in D$ area.

Let us define the first implementation of $f_r$ as a biquadratic regression equation in two variables $(x, y)$ (3), to find its $a \in A$ coefficients by the gradient descent method, we find the minimum of the residual function (4), (5).

$$f_r(x, y) = a_0 + a_1 x + a_2 y + a_3 xy + a_4 x^2 + a_5 y^2 + a_6 x^2 y + a_7 xy^2 + a_8 x^2 y^2, \tag{3}$$

$$F(a) = \sum_{i=0}^{n} (f(x_i, y_i) - f_m(x_i, y_i))^2, \tag{4}$$

$$F(a) \to \min. \tag{5}$$

As seen in Fig. 2 b, the value of the biquadratic regression equation, despite satisfying the condition of the minimum discrepancy (4), in some areas in the boundary of the domain of definition significantly goes beyond the oceanographic range. To solve this problem, we will implement a flexible threshold smoothing of the function when it goes beyond the boundary values of the original function in the form (5).

In this case, $f_l$ it is formed as a functional from $f_r$, when the value $f_r$ goes beyond $\max f(x, y)$ or $\min f(x, y)$, the function is compressed to its nearest boundary, the surface graph of which is shown in (Fig. 2 c).

$$f_l(x, y) = \begin{cases} \max f(x, y) + \dfrac{1}{f_r(x, y) - \max f(x, y) + \beta}, & \text{если} \quad f_r(x, y) \geq \max f(x, y) \\ \min f(x, y) - \dfrac{1}{\min f(x, y) - f_r(x, y) + \beta}, & \text{если} \quad f_r(x, y) \leq \min f(x, y) \\ f_r(x, y), & \text{иначе} \end{cases} \tag{5}$$

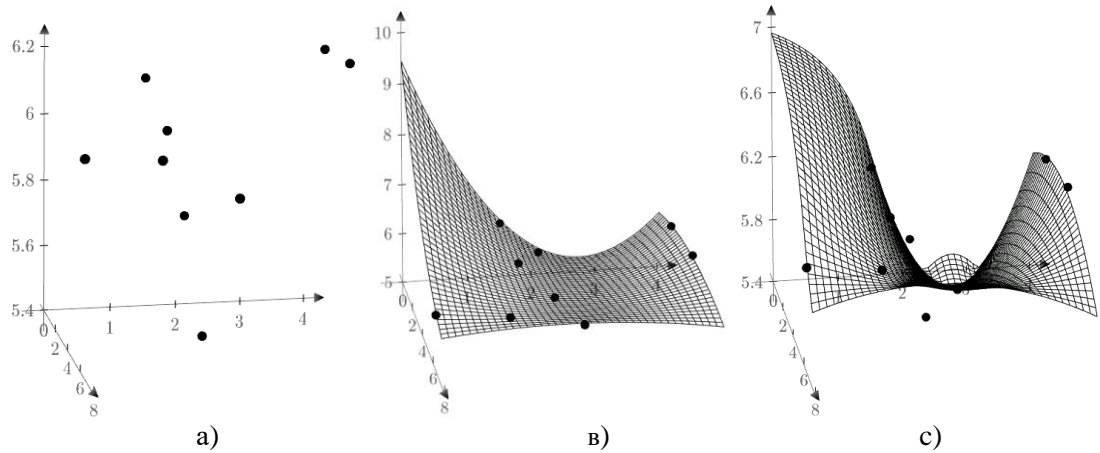

**Fig. 2.** a) a scatter plot of the original data; b) plot of the surface of the regression model, given as a biquadratic function $f_r$; c) plot of the surface of the regression model given as a biquadratic function by flexible threshold smoothing $f_l$

Let us construct a gradient field obtained by the $f_l$ regression model and, in accordance with (1) at each node of this field, calculate the coverage area of the meter. For clarity of the obtained result, we will project the resulting field onto a map of the area (Fig. 3). In this example, as an example of the resolution, measuring channels equal to 0,001 ml/l are used, for example, the Seabird SBE 43 dissolved oxygen sensor [18].

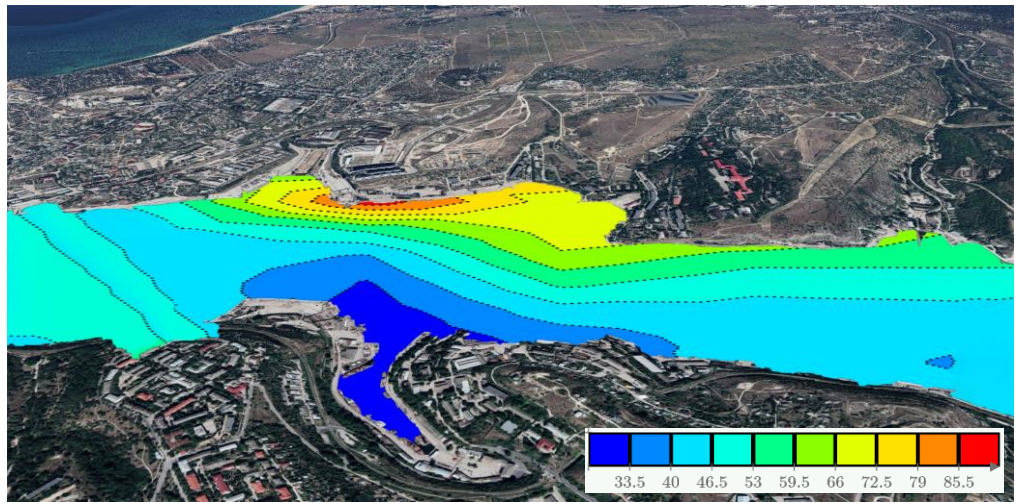

**Fig. 3.** Heat map of the optimal intervals for measurements of dissolved oxygen in the surface layer of the Bay of Sevastopol, m

The resulting coverage area displays at what interval it is optimal to take measurements in each specific area.

**Conclusion**. The proposed conceptual model of a system for monitoring the main parameters of the aquatic environment, built on the basis of a distributed system of small-sized intelligent autonomous surface and underwater robots, differs from existing prototypes in that it allows automated comprehensive monitoring of the main parameters of the environment throughout the entire field of observation.

The operation of the algorithm for constructing maps of the optimal intervals for measuring the parameters of the aquatic environment is illustrated by the example of constructing a heat map for measuring the dissolved oxygen in the surface layer of the bay of Sevastopol. This can serve as a confirmation of the work of the intelligent decision support system when organizing the optimal process of station location.

The constructed maps of optimal measurement intervals can serve as input material for the development of IASR control systems when solving problems of time optimization, energy saving and the number of stations.

The proposed approach to solving the problem of monitoring the aquatic environment using a decentralized IASR network has the following advantages: scalability, flexibility, promptness of deployment and collapse, self-organization, the ability to create a wide field of view by changing the number of IASR.

The proposed model was developed for use in adoption support systems for continuous automated control of the IASR monitoring process without the participation of operators, which, due to the systematic data collection, allows one to obtain new scientific results.

*The study was funded by Russian Foundation for Basic Research and Government of the Sevastopol according to the research project No. 18-48-920018\18.*

## REFERENCES

1. *Venkatesan R.* Observing the Oceans in Real Time. Springer International Publishing, 2018.

2. *Marc L.M.* Instrumentation and Metrology in Oceanography, ISBN: 978-1-848-21379-1, Sep 2012, Wiley-ISTE, 393 p.

3. *Bellingham J.B.* New oceanographic uses of autonomous underwater vehicles // Mar. Technol. Soc. J., 1997. Vol. 31, no. 3. P. 34–47.

4. *Curtin T.B., Bellingham J.G., Catipovic J.* Autonomous oceanographic sampling networks // Oceanography. 1993. Vol. 6 (3). P. 86–94.

5. *Minaev D.D.* Principles of building a regional automated information system for environmental monitoring of sea areas using autonomous technical means and robotic systems // Underwater Research and Robotics. 2011. Vol. 2 (12). P. 64–68.

6. *Emelyanova V.P., Lobchenko E.E.* RD 52.24.643-2002. A method for a comprehensive assessment of the degree of pollution of surface waters by hydrochemical indicators. Depon. Moscow, 2004. 20 p.

7. *Mezentseva I.V., Malchenko Yu.A.* An integrated approach to the organization of monitoring sea water pollution in the coastal waters of Sevastopol // Proceedings of the State Oceanographic Institute. 2015. No. 216. P. 326–339.

8. *Trusevich V.V., Gaiskii P.V., Kuz'min K.A.* Automatic biomonitoring of aqueous media based on the response of bivalves // Physical Oceanography. 2010. Vol. 20. № 3. C. 231–238.

9. *Gaiskii P.V., Trusevich V.V., Zaburdaev V.I.* Automatic bioelectronic complex for early detection of toxic pollution in fresh and sea waters // Marine Hydrophysical Journal. 2014. No. 2. P. 44–53.

10. *Benjamin M.R., Schmidt H., Newman P.M.* Nested autonomy for unmanned marine vehicles with MOOS-IvP // J. Field Robot. 2010. 27 (6). P. 834–875. DOI: 10.1002/rob.20370

11. *Schmid, H., Benjamin M.R., Petillo S.M.* Nested autonomy for distributed ocean sensing // Springer Handbook of Ocean Engineering / eds. M.R. Dhanak, N.I. Xiros. New York: Springer, 2016. P. 459–480.

12. *Mahmoudian N., Woolsey C.* Underwater glider motion control, in 2008 IEEE Conference on Decision and Control, 552–557. doi: 10.1109/CDC.2008.4739432

13. *Leonard J.J., Bahr A.* Autonomous underwater vehicle navigation // Springer Handbook of Ocean Engineering. New York: Springer. 2016. P. 341–358.

14. *Kostenko V.V., Lvov O.Yu.* Combined communication and navigation system of an autonomous underwater robot with a float module // Underwater Research and Robotics. 2017. P. 31–43.

15. *Heidemann J., Stojanovic M., Zorzi M.* Underwater sensor networks: applications, ad-

vances and challenges // Phil. Trans. R. Soc. A. 2012. T. 370, № 1958. P. 158–175.

16. *Lermusiaux P.F., Subramani D.N., Lin J.* A future for intelligent autonomous ocean observing systems // Journal of Marine Research. 2017. 75 (6). P. 765–813.

17. *Konovalov S.K., Romanov A.S., Moiseenko O.G.* Atlas of Oceanographic Characteristics of the Sevastopol Bay. Sevastopol: "ECOSY-HYDROPHYSICS", 2010. 320 p.

18. *Seabird* SBE 43 and SBE 43F individually calibrated, high-accuracy oxygen sensor to assist in critical hypoxia and ocean stoichiometric oxygen chemistry research on a variety of profiling and moored platforms Datasheet (Available from http://seabird.com/oxygen-sensors/sbe-43-dissolved-oxygen-sensor)